\documentclass[aps,onecolumn,prd,showpacs,nofootinbib]{revtex4}

\def\be{\begin{equation}}
\def\ee{\end{equation}}
\def\ba{\begin{eqnarray}}
\def\ea{\end{eqnarray}}
\def\bff{\begin{figure}}
\def\eff{\end{figure}}

\usepackage{graphicx}
\usepackage{epsf}
\usepackage{xcolor}
\usepackage{graphics}
\usepackage{amsmath}
\usepackage{dcolumn}
\usepackage{bm}
\usepackage{amssymb}
\usepackage{latexsym}
\usepackage{float}
\usepackage{url}

\usepackage[dvipdfm]{hyperref}

\begin{document}

\title{\Large{Inhomogeneous reheating scenario with DBI fields} \vspace{6mm}}

\author{\large{Sheng Li}}
\email{sl277@sussex.ac.uk}
%\ead{sl277@sussex.ac.uk}
\affiliation{%\large{
$^1$
Astronomy Centre, University of Sussex, Brighton\\
BN1 9QH, United Kingdom
}

\begin{abstract}
\vspace{10mm}
\begin{center} \textbf{Abstract} \end{center}
We discuss a new mechanism which can be responsible for the origin of the primordial perturbation
in inflationary models, the inhomogeneous DBI reheating scenario. Light DBI fields fluctuate during inflation,
and finally create the density perturbations through modulation of the inflation decay rate. In this note, we investigate the curvature perturbation and its non-Gaussianity from this new mechanism. Presenting generalized expressions for them, we show that the curvature perturbation not only depends on the particular process of decay but is also dependent on the sound speed $c_s$ from the DBI action. More interestingly we find that the non-Gaussianity parameter $f_{NL}$ is independent of $c_s$. As an application we exemplify some decay processes which give a viable and detectable non-Gaussianity. Finally we find a possible connection between our model and the DBI-Curvaton mechanism.

\end{abstract}
%\pacs{98.80.Cq}
\maketitle

\vskip 10mm
\section{Introduction}
To account for the density perturbation which seeds the structure of the observed universe, the inflationary paradigm\cite{KT}
is a promising candidate. In this picture, the universe went through an accelerated expansion in the very early period. This
scenario predicts that the inflaton field $\phi$ rolls downs its potential with quantum fluctuations superimposed which lead to density perturbations. However it is important to investigate alternatives to this simple scenario.

Recently more and more evidence from observations of cosmic microwave background (CMB) anisotropies favors that
the primordial density fluctuations (PDF) are almost Gaussian, Scale invariant, and adiabatic \cite{Spergel,Komatsu:2010fb}. And, according to many works \cite{Lyth:2001nq,Moroi:2001ct,Dimopoulos:2002kt,Lyth:2002my,Lyth:2005ab,Lyth:2005fi,Dvali:2003em,Kofman:2003nx,Dvali:2003ar,Zaldarriaga:2003my,Matarrese:2003tk,Suyama:2007bg,Battefeld:2007st,Ichikawa:2008ne},
 light scalar fields merit attention to investigate their role as candidates for explaining the origin of the PDF.
 Such light scalar fields generally exist in extensions of the standard model of particle physics, which motivates some alternatives. One is the curvaton scenario
\cite{Lyth:2001nq,Moroi:2001ct,Dimopoulos:2002kt,Lyth:2002my,Lyth:2005ab,Lyth:2005fi}, in which the final curvature perturbations are produced from an initial isocurvature perturbation associated with the quantum fluctuations of a light scalar field other than the inflaton, the curvaton, whose energy density is negligible during inflation. The curvaton
isocurvature perturbations are transformed into adiabatic ones when the curvaton decays into radiation well after the end of inflation. Another is the inhomogeneous/modulated reheating scenario
\cite{Dvali:2003ar,Dvali:2003em,Zaldarriaga:2003my,Kofman:2003nx,Matarrese:2003tk,Vernizzi:2003vs}. This supposes that the decay rate $\Gamma$ of the inflaton varied in space due to a dependence on a light field, and density perturbations would be generated during reheating independently of those generated by the standard inflationary mechanism. In these two scenarios the light scalar fields, not the inflaton, are responsible for the primarily density perturbations. The inflaton just serves to drive and end inflation, and under this assumption the constraints on inflation are considerably lessened. By introducing these light scalar fields one also finds that primordial non-Gaussianity can be very large compared to the single inflaton models. In curvaton models the non-linear parameter $f_{NL} \sim 5/4r$ where $r$ is a small coefficient \cite{Lyth:2002my,Lyth:2005ab,Lyth:2005fi}; and inhomogeneous/modulated reheating models suggest $f_{NL} \sim O(1)$ \cite{Dvali:2003ar,Zaldarriaga:2003my,Matarrese:2003tk} or larger given particular decay processes, see Refs \cite{Suyama:2007bg,Battefeld:2007st,Ichikawa:2008ne,Kohri:2009ac}.

Cosmological modellers have paid attention recently to the DBI field as an alternative to a canonical field. Some earlier works, Refs.\cite{Silverstein:2003hf,Alishahiha:2004eh,Chen:2006nt,Easson:2007dh,Easson:2007fz,Gmeiner:2007uw,Arroja:2008yy}, have discussed cosmology using a DBI field, and with this interesting source some results are different from those obtained in Refs.
\cite{Huang:2007hh,Langlois:2008mn,Langlois:2008wt,Langlois:2008qf,Guo:2008sz}. So far only the ordinary light scalar fields were discussed in inhomogeneous reheating, therefore in this note, we focus on the inhomogeneous(modulated) reheating models specifically involving DBI fields. As a toy model to reheat the universe, we assume that the DBI fields can dominate the decay rate.
With the variation of the decay rate, the fluctuations can be generated after inflation.

\section{Basic Mechanism}
To reheat the universe, in the scenario of Ref.\cite{Dvali:2003em} the inflaton couples with ordinary particles. When the inflaton decays, the decay rate has the form $\Gamma \sim \lambda^2m$, where $\lambda$
is a stochastic variable and $m$ is the mass of the inflaton. Assuming $\lambda$ is a
function of the scalar fields in the theory, contrary to the standard scenario, the fluctuations
in this new scenario are determined by the fluctuations of the decay rate $\Gamma$ and not the
inflaton field $\phi$. This means that when two places have a different decay rate, the cosmological
evolution in these regions will undergo different processes and eventually result in
density perturbations when reheating finishes. Finally, the density perturbations are
\begin{equation}
\frac{\delta\epsilon_{rad}}{\epsilon_{rad}} \propto
\frac{\delta\lambda}{\lambda} \propto \frac{\delta\Gamma}{\Gamma}
\end{equation}
where we find the fluctuations in $\Gamma$ are transferred into density perturbations. More
detailed discussions can be found in Ref.\cite{Dvali:2003em}.

In our model, which is motivated by Refs.\cite{Dvali:2003em} and \cite{Li:2008fm}, we consider the decay rate to be determined by a DBI field. According to these two papers, we know the light fields are expected to provide a considerable perturbation, and in Ref.\cite{Li:2008fm} we can see that light DBI-fields lead to a different and larger curvature perturbation. We also get a factor $f_{NL}^{\rm equi}$ which is determined by the sound speed $c_s$, which gives a sizable non-Gaussianity.

Expecting to get further understanding of origin of perturbations after inflation, we propose the decay rate to be determined by the DBI field in the scenario of inhomogeneous reheating. To investigate the relationship between the decay rate and the light fields, we generalize $\Gamma$ to a form $\Gamma_\phi(\sigma) \equiv f(\sigma)$ where the subindex $\phi$ denotes the inflation era, and $\sigma$ is the light DBI field. In general $\Gamma$ varies with the space-time location.

For further discussions, it is necessary to quote some results from Ref.\cite{Li:2008fm}, which described the DBI curvaton scnario and which gives the fluctuations of the light scalar field $\sigma$, curvature perturbation $\zeta$ and its non-Gaussianity parameter $f_{NL}$.
\ba
\delta\sigma &=& \sqrt{c_s} \frac {H_*}{2\pi} \label{dbi-sig}\\
\zeta &\sim& \frac{h_{,\sigma}}{h} \frac{1}{c_s^{3/2}}\frac{H_*}{12\pi}  \label{dbi-zeta}\\
f_{NL}&\simeq& \frac{5}{2}\left[\frac{3(1-c_s^2)^3}{(1-3c_s^2)^2}+ \frac{4h}{h_{,\sigma}^2}\left(\frac{h_{,\sigma\sigma}}{2}-\frac{h_{,\sigma}^2}{h}\right)\frac{c_s^4(1-c_s^2)}{1-3c_s^2}
\right] \label{dbi-fnl}
\ea
Here the sound speed $c_s$ is defined as $c_s = \sqrt{1-2 h(\sigma) X}$, $X\equiv-\frac{1}{2}g^{\mu\nu}\partial_{\mu}\sigma\partial_{\nu}\sigma -\frac{1}{2}g^{\mu\nu}\partial_{\mu}\phi\partial_{\nu}\phi$
in which $\phi$ plays the role of an inflaton field while $\sigma$ is the DBI field, and $h(\sigma)$ is the warping factor. We stress the amplitude of the curvature perturbation, and more detail can be found in Ref.\cite{Li:2008fm}.

Before discussing the parameters which describe the properties of cosmological evolution, such as curvature perturbations and the non-linear parameter $f_{NL}$, we can expand the decay rate to second order
\ba
\label{gamma}
\Gamma &=& \Gamma_* + \delta^{(1)}\Gamma + \frac{1}{2}\delta^{(2)}\Gamma  \nonumber \\
&=& f(\sigma_*) + f'(\sigma_*) \delta\sigma + \frac{1}{2} f''(\sigma_*)(\delta\sigma)^2
\ea
Here $'$ denotes the derivative with respective to the DBI-field $\sigma$, and $H_*$ and $\Gamma_*$
represent the Hubble parameter during inflation and the homogeneous value of the decay rate respectively.
%From above we can easily write
%\begin{eqnarray}
%\frac {\delta\Gamma}{\Gamma_*} = \frac{f'(\sigma_*)}{f(\sigma_*)}\delta\sigma +
%\frac{1}{2}\frac{f''(\sigma_*)}{f(\sigma_*)}(\delta\sigma)^2
%\end{eqnarray}

Now, let's look at the general representation of the curvature perturbation. Refs.\cite{Dvali:2003ar,Zaldarriaga:2003my,Maldacena:2002vr,Scoccimarro:2003wn,N.Bartolo:2004uv} show that the curvature perturbation $\zeta$ can be expressed by the fluctuations of the decay rate, $\delta\Gamma/\Gamma_*$
\be\label{nzeta}
\zeta = -\alpha \log \frac{\Gamma}{\Gamma_*} = -\alpha \frac{\delta\Gamma}{\Gamma_*}
\ee
where the coefficient $\alpha$ can be determined by the quantity $\Gamma_* / H_*$. Ref.\cite{Dvali:2003em} proved that during the inflation era, in the limit $\Gamma_*/H_* \rightarrow 0$, $\alpha = 1/6$. Substituting Eq.(\ref{gamma}) into Eq.(\ref{nzeta}), we can get the curvature perturbation in the linear approximation
\be
\label{dbi-inho}
\zeta = -\sqrt{c_s} \frac{f'(\sigma_*)}{f(\sigma_*)} \frac{H_*}{12\pi}
\ee
Note that the final curvature perturbation in this scenario depends not only on the sound speed $c_s$ but also on the ratio $f'/f$ which comes from the details of the decay process. They will be discussed in detail in the following sections.

\section{Non-Gaussianity}

Let's now focus on the degree of non-Gaussianity. During inflation, both $\phi$ and $\sigma$ are slowly rolling. It is known that the non-Gaussianity of $\zeta$ coming from the intrinsic non-Gaussianities of $\delta \phi_*$ and $\delta \sigma_*$ is far below the observational sensitivity. Hence we can treat $\delta\phi_*$ and $\delta \sigma_*$ as uncorrelated Gaussian random fields with the same amplitude, see Ref. \cite{Ichikawa:2008ne} and references therein.

As we know, the standard inflationary scenario predicts the degree of non-Gaussianity with $f_{NL} \sim tilt$ of the perturbation spectrum \cite{Maldacena:2002vr,Scoccimarro:2003wn,N.Bartolo:2004uv}. By contrast, in the curvaton scenario, significant non-Gaussianities are easily produced because the curvaton density is proportional to the square of the curvaton \cite{Lyth:2005fi}, giving $f_{NL} \sim 5/4r $. Some recent works related to DBI fields also give larger positive non-Gaussianities \cite{Q.Huang:DBIs,Y.Cai:DBIs} and some related topics have been discussed in the isocurvaton scenario \cite{M.Li:DBIs}. Non-Gaussianity in models where density perturbations are produced by spatial fluctuations in the decay rate of the inflaton have been discussed in Refs. \cite{Dvali:2003ar,Zaldarriaga:2003my}. They show that $\it{f_{NL} \sim few}$, which is larger than that coming from the inflation model, and possibly accessible to future observations.

The current observational data $-10< f_{NL}^{local} < 74$ \cite{Komatsu:2010fb,Larson:2010gs} permits the primordial non-Gaussianity to  be large. In the case of a low energy scale, however, we can safely ignore the contribution of inflaton $\phi_*$ to the non-Gaussianities, and we just consider the rest of the light scalar fields which are mainly responsible for the generation of non-Gaussianities.

To calculate the local form of non-Gaussianity, for simplicity we assume there is only one DBI field. With the definition of $f_{NL}$ appearing in Ref. \cite{Maldacena:2002vr,Scoccimarro:2003wn,N.Bartolo:2004uv} (and references therein) it corresponds to
\ba
\label{inhz}
\zeta &=& \zeta_g - \frac{3}{5}f_{NL}(\zeta_g)^2
\ea
where the coefficient $f_{NL}$ \footnote{Note: In Refs. \cite{Wands:2010af} and most other papers, the formula $\zeta = \zeta_g + 3/5 f_{NL}(\zeta_g)^2$ is used, but here we take the opposite sign for $f_{NL}$. The conventions can be linked by taking $f_{NL} = -f_{NL}$.} represents the non-Gaussianity parameter of the curvature perturbation.
Looking back to Eq.(\ref{dbi-inho}) up to second order, it gives
\ba
\label{expinhz}
\zeta &=& -\frac{1}{6}
\left\{\frac{f'(\sigma_*)}{f(\sigma_*)}\delta\sigma + \frac{1}{2} \left[\frac{f''(\sigma_*)}{f(\sigma_*)}
-\left(\frac{f'(\sigma_*)}{f(\sigma_*)}\right)^2 \right] (\delta\sigma)^2 \right\}
\ea
Comparing with Eq.(\ref{expinhz}) and Eq.(\ref{inhz}), we get the parameter as the function of $f(\sigma)$
and its first and second order derivatives,
\ba
\label{in-fnl}
f_{NL} &=& 5 \frac{ \frac{f''(\sigma_*)}{f(\sigma_*)} - \left(\frac{f'(\sigma_*)}{f(\sigma_*)}\right)^2 } {\left(\frac{f'(\sigma_*)}{f(\sigma_*)}\right)^2}= 5 \left( \frac{f(\sigma_*)f''(\sigma_*)}{f'^2(\sigma_*)} - 1 \right)
\ea
Similar results are presented in Ref.\cite{Ichikawa:2008ne}. We find that the precise value of $f_{NL}$ depends on specific models, but very large non-Gaussianity $|f_{NL}| \gg 1$ is obtained when $\left|\frac{f(\sigma_*)f''(\sigma_*)}{f'^2(\sigma_*)}\right| \gg 1$ is satisfied. Detailed discussions are given in the following section.

\vskip 1cm
\section{Discussion}
\subsection{Observable Parameters}
As mentioned above, we have found some results which differ from previous works. From Eq.(\ref{dbi-inho}), we can see that the curvature perturbation becomes ${\cal{P}}^{1/2}_{\zeta} \sim {\frac{\sqrt{c_s}}{12\pi}}\frac{H_*}{M_*}$; here $\rm M_*$ is the mass scale during reheating. In the inflationary background, gravitational waves can be predicted as the tensor fluctuation ${\cal{P}}_{T}^{1/2} \sim {\frac{1}{\pi} \frac{H_*}{M_{\rm pl}}}$.

\textbf{(1)} At first glance, due to the existence of the sound speed the primordial curvature perturbation, $\zeta$ in Eq.(\ref{dbi-inho}), can be suppressed by the term $\sqrt{c_s}$, but we also note that the detailed decay process plays a significant role in determining the final value of the curvature perturbation.

\textbf{(2)} Considering the ratio $r$ of tensor to scalar, here in our model it is given by
\be
r = \frac{{\cal P}_{T}}{{\cal P}_{\zeta} }\sim 10^2 \frac{1}{c_s} \frac{f^2}{M_{pl}^2 f'^2}
\ee
Like any linear approximation, in expanding we assume that the decay process has the form of $f(\sigma) \propto (1 + \mathrm{\vartheta \sigma/M} + ... )$ (similarly see Refs.\cite{Dvali:2003em,Ichikawa:2008ne}), where $\vartheta$ is small, and $\rm M$ (by ignoring subindex $*$) is a dimensional parameter representing the mass scale for reheating after inflation ending, and $m_\sigma < M < M_{pl}$. If we use $r < 0.36 (95\% \rm{CL})$ from 7-year WMAP limit in the Ref.\cite{Komatsu:2010fb}, then we should impose $f'/f \sim \vartheta/M_*$, so that $M_* \sim \vartheta \sqrt{c_s} M_{\rm pl} \times 10^{-3}$. This means that for very small $c_s$, we can get the $m_\sigma \ll M_{\rm pl}$ which is assumed in the beginning of our consideration.

\textbf{(3)} Contrary to the standard inhomogeneous reheating scenario of Ref.\cite{Dvali:2003em}, in our model the created
fluctuations are of order $\sqrt{c_s} H_* \frac{f'}{f} \sim \vartheta \sqrt{c_s} \frac{H_*}{M_*}$. The relation between the slope of the power spectrum of density perturbations and the inflaton potential in various scenarios is,
\ba
n-1 &=& \frac{d \ln{ \Big((f'/f)^2c_sH^2}\Big) }{d \ln{a}}\quad \quad \quad \mbox{in our model} \nonumber \\
n-1 &=& \frac{d \ln{H^2}}{d \ln {a}}\quad \quad \quad \quad \quad \quad \quad \quad \mbox{standard inhomogeneous reheating} \nonumber \\
n-1 &=& \frac{d \ln{H^2/\epsilon}}{d \ln{a}}\quad \quad \quad \quad \quad \quad \quad  \mbox{standard inflation scenario}
\ea
Our result is different from both Ref.\cite{Dvali:2003em} and the standard inflationary scenario.

Discussing the spectral index explores to what degree the decay process and the warp space contribute to it. With $c_s = \sqrt{1-2Xh(\sigma)}$ and its derivative with respective to DBI field $\rm \sigma$, i.e. $c_s^\prime = \frac{(c_s^2 - 1)}{2c_s} \frac{h'(\sigma)}{h(\sigma)}$, then we obtain
\ba
\label{h-van}
n-1 = \frac{d \ln{H_*^2}}{d\ln a} + \left[ \frac{c_s^2 - 1}{2c_s^2} \frac{h'(\sigma)}{h(\sigma)} + 2 \frac{f'(\sigma)}{f(\sigma)} \left(\frac{f(\sigma)f''(\sigma)}{f'^2(\sigma)} - 1\right)  \right] \frac{d\sigma}{d\ln{a}}
\ea
where parameter $h(\sigma)$ represents the warped throat and $f(\sigma)$ indicates the decay rate. For simplicity for estimating the contribution to power spectra, we set all $h'(\sigma)/h(\sigma)$ and $f'(\sigma)/f(\sigma)$ as proportional to $1/\sigma$ for easy estimation. By doing this, we can combine the second term as
\ba
n-1 &\sim& \frac{d \ln{H_*^2}}{d\ln a} + \left[ \frac{c_s^2 - 1}{2c_s^2} +
2 \left(\frac{f(\sigma)f''(\sigma)}{f'^2(\sigma)} - 1\right)  \right] \frac{d\ln{\sigma}}{d\ln{a}} \nonumber\\
 &\sim& \frac{d \ln{H_*^2}}{d\ln a} + \left( \frac{c_s^2 - 1}{2c_s^2} +
\frac{2}{5} f_{NL}  \right) \frac{d\ln{\sigma}}{d\ln{a}}
\ea

\textbf{(A):} if $c_s \simeq 1$, the decay rate surpasses the contribution from the warped throat. From Eq.(\ref{h-van}), the effect of the warp throat is reduced. However, in view of the Lagrangian, the system is just reduced to a canonical one, and can be rewritten as a usual dynamic system with $\mathcal{L} = \mathcal{L} {(X,\sigma)} = {-\frac{1}{2}{\partial_{\mu}{\sigma}\partial^{\mu}{\sigma}} - V(\sigma)}$.

\textbf{(B):} if $c_s \ll 1$, it is hard to say which contribution affects the whole term more.
Because of the warped factor $h$, the sound speed $c_s$ and the unknown decay process $f(\sigma)$, the model possesses a lot of freedom, but reheating from a warped throat is worthy of investigation in future work.

\subsection{Some Specific Examples}
Interestingly, from Eq.(\ref{in-fnl}), the non-Gaussianity is independent of the sound speed in our model even though the curvature perturbation depends on it. This distinguishes from the previous works in DBI field scenario \cite{Huang:2007hh,Langlois:2008mn,Langlois:2008qf,Li:2008fm} where the primordial fluctuations and non-Gaussianities are both enhanced by a low sound speed. Here the non-Gaussianities only depend on the specific decay process.

We wish to know whether a particular decay process exists to generate a large curvature
perturbation and also non-Gaussianities. Let's consider some specific toy models able to
produce detectable density perturbations and large non-Gaussianity.

\begin{figure}[!ht]
%\setlength{\parskip}{-5pt}
%Horizonal Multifigures
\setlength{\abovecaptionskip}{0pt} %\flushleft
%\centering
%\includegraphics[width=0.3\textwidth]{toy-log-pert2}
%\begin{tabular}{cc}
%\epsfig{file=
\begin{minipage}[b]{0.48\linewidth} % A minipage that covers half the page
\includegraphics[width=2.8in]{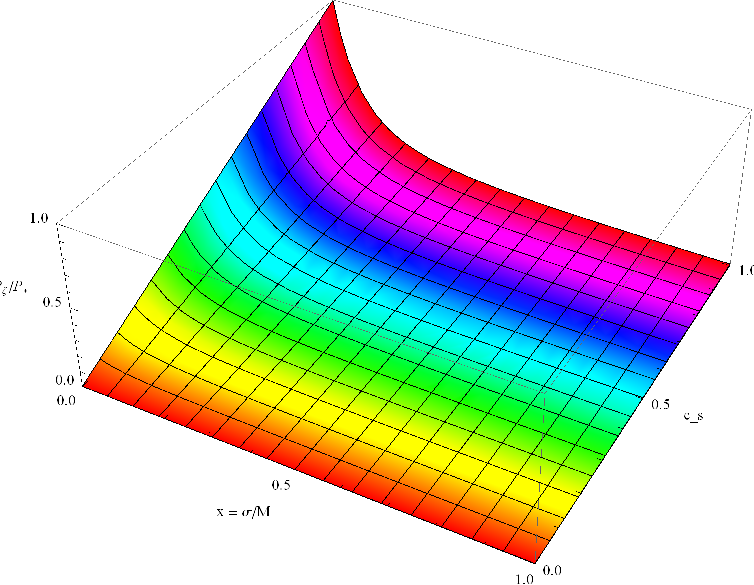}
\hspace{0.4cm}
\label{fig:zeta}
\end{minipage}
\begin{minipage}[b]{0.48\linewidth} % A minipage that covers half the page
%{width=0.33\linewidth,clip=1cm} &
\includegraphics[width=2.8in]{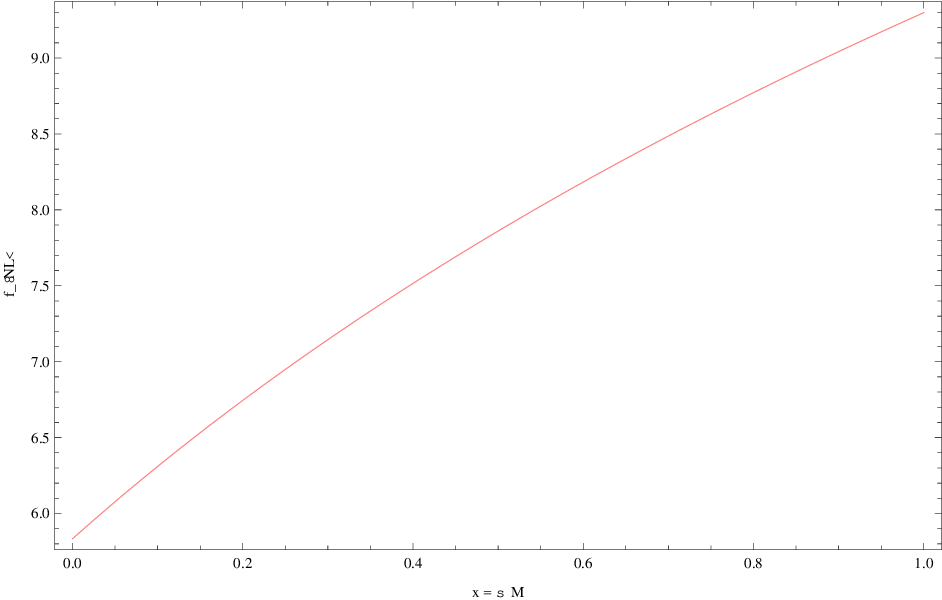}%{width=0.33\linewidth,clip=1cm}
\label{fig:fnl}
%\end{tabular}
\end{minipage}
\caption{(Left) Plot of the ratio $\mathcal \Re = P_\zeta / P_*$, where ${P_*}$ indicates the scale-invariant power spectrum $P_* = \left(\frac{H_*}{2\pi}\right)^2$.
(Right) non-Gaussianity of perturbation. The plots show that the amplitude of $f_{NL}$ is of O(1) , even though $\sigma \ll { M}$. There, the mass scale for $M$ is lower than inflaton scale ${H_*}$.}\label{fig:Curvature-fnl}
\end{figure}
\textbf{(1)} We assume that the decay process is described by $f(\sigma) \sim \Big(1 + 6 \log \left(1 + \sigma/{\rm M} \right) \Big)^{-1}$. Here $\rm M$ is a mass scale during reheating, which is less than $\rm H_*$.

In this case with Fig.1, we find $\zeta \sim \frac{\sqrt{c_s}}{(1 + 6\log (1+\sigma/{\rm M}))(1+\sigma/ {\rm M})}$ and $f_{NL} = \frac{35}{6} + 5\log(1+\sigma/ {\rm M})$. From the plot on the left hand side, when the speed of DBI field $\sigma$ reaches its relativistic limit, which means the sound speed $c_s$ approaches zero, we get the curvature perturbation converted from the fluctuation of $\sigma$ to be zero. If the sound speed is approaching $1$, meaning corresponding to a canonical system, the curvature perturbation can be of order $P_*$. However the ratio $\mathcal \Re$ is less than $1$, from the left plot, generally the curvature perturbation of the DBI field through decaying process is suppressed by $c_s$, even if the DBI field has a large mass scale.

We can see the non-Gaussianity from the plot on the right hand side. It is independent of the sound speed $c_s$. It is possible to obtain a large non-Gaussianity, being of O(10), if we place suitable coefficients in the decay rate.

\textbf{(2)} If we take for example the form of $\left(1 - \lambda  \sigma / M \right)^{-1/n}$, we could get $f_{NL} \sim O(1 \sim 10)$ or more. Here $\lambda$ is just a sign which is used to determine the curvature perturbation together with the power $n$, while $M$ is a mass scale during reheating.
\begin{figure}[!ht]
%\begin{tabular}{cc}
\begin{minipage}[b]{0.48\linewidth} % A minipage that covers half the page
%\centering
\includegraphics[width=2.8in]{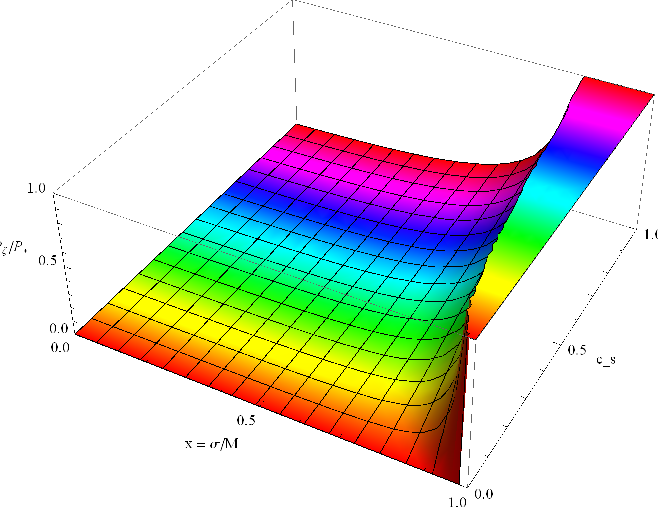}%\linewidth,clip=1cm} &
%\caption{Plot for the case of $n>0$.}
\label{fig:n}
\end{minipage}
\begin{minipage}[b]{0.48\linewidth} % A minipage that covers half the page
\hspace{0.4cm}
\includegraphics[width=2.8in]{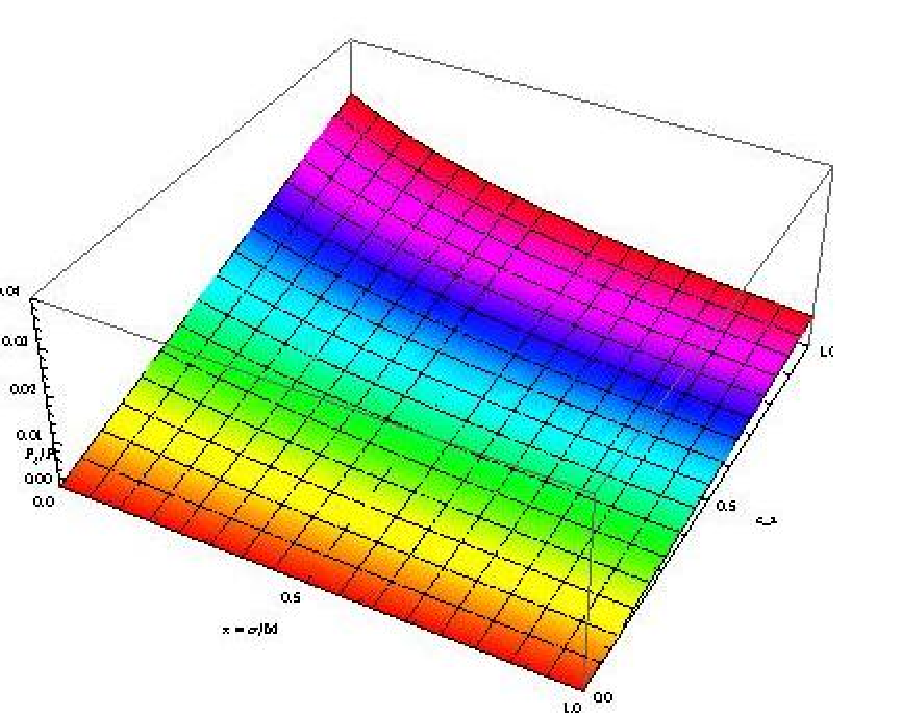}%\linewidth,clip=1cm}
%\end{tabular}
%\caption{Plot for the case of $n<0$.}
\label{fig:-n}
\end{minipage}
\caption{ Different shapes of the ratio $\mathcal \Re = \frac{P_\zeta}{P_*}$ with $n>0$ (Left) and $n<0$ (Right). Here we place $n=1$. When $\left|n\right|$ is larger than one, the perturbation will be more flat.}
\label{fig:Curvature-fnl}
\end{figure}

In Fig. 2, we notice the curvature perturbations have different shapes in case with different $n$ and the sound speed $c_s$ overall. The figures show that the perturbations are smoothed in different parametric regions. The non-Gaussianity $f_{NL}$ of these types of perturbation has the same shape, with $f_{NL} \equiv 5n$ regardless of the sign of $n$, even given the different $c_s$.

If $c_s \sim 0$, the curvature perturbation approaches zero too. When $c_s$ gets larger, we can find the possibility of generating a detectable perturbation when $n>0$ (in the left figure). In contrary, when $n<0$ (in the right figure), it might not easy to detect the curvature perturbation which is converted from the DBI field during the decay period.

\section{Summary and Outlook}

In this note we proposed a variant on the inhomogeneous reheating model, in which the decay rate of the inflaton is determined by DBI fields. During inflation, the light DBI fields  fluctuate and dominate the final density perturbations through the modulated inflaton decay rate. We gave the general expression for the curvature perturbations and non-Gaussianity for this model. The decay rate of the inflaton played an important role in determining their final value. By presenting the generalized expressions for them we showed that the curvature perturbation $\zeta \sim \sqrt{c_s}\frac{f'}{f} $ depends not only on the particular process of decay but also on the sound speed $c_s$ from the DBI action. Apparently it is suppressed by sound speed, but due to the uncertain process of decay there will be other possible outcomes. Moreover the non-Gaussianity in our model is independent of $c_s$ no matter that $\zeta$ depends on it. As discussed in this article, we find the non-linear parameter $f_{NL}$ is independent of the sound speed $c_s$ and large non-Gaussianity can be obtained if $\left|f''f/f'^2\right| \gg 1$ is satisfied. As an application we exemplify some kinds of decay process leading to a detectable non-Gaussianity, $f_{NL}$ of O($1\sim10$) which is compatible with observational data.

As a byproduct and outlook, comparing Eq.(\ref{dbi-inho}) with Eq.(\ref{dbi-zeta}), we can rewrite them as $\zeta_{\rm {inhomo}}^{\rm {DBI}} \sim -\vartheta_{i} \label{dbi-zinhomo1}$
and $\zeta_{\rm {curvaton}}^{\rm {DBI}} \sim \frac{1}{c_s^2} \vartheta_{c}  \label{dbi-zeq1}$
where we define two factors respectively for the above two quantities
$\vartheta_{i} \equiv \sqrt{c_s} \frac{f'(\sigma_*)}{f(\sigma_*)}$ and $\vartheta_{c} \equiv \sqrt{c_s} \frac{h'(\sigma_*)}{h(\sigma_*)}$.

If we consider a tachyon condensation, as discussed in Ref.\cite{Li:2008fm}, where the DBI-Curvaton has this kind of exit mechanism rather than oscillation as in usual curvaton model, then the universe can be instantly reheated. However, we presume that in a warped throat, in which a space-time has a strong compactification, that the motion of DBI fields $\sigma$ or the D-brane itself could be assumed to decay once the inflation ends. Due to the fields having a velocity limit in a D-brane, which could be interpreted by the sound speed $c_s$ in DBI action, it is interesting to bridge the decay process and the warp throat with the sound speed, namely $f(\sigma) \simeq h(\sigma)c_s^2$. Then we may find that the DBI-Curvaton is a special case in the context of this model if there should exist a decay process in the form of $f(\sigma) \simeq h(\sigma)c_s^2$ in a warped throat. To investigate the inhomogeneous reheating mechanism concerning a warped throat is a valuable direction for future work, although many works have been done in the context of a throat \cite{Barnaby:2004gg,Chen:2005ad,Chialva:2005zy,Kecskemeti:2006cg,Becker:2007ui,Alabidi:2010ba,Kobayashi:2009cm,Zhang:2009gw}.

\vskip 5mm
{\it \textbf{Acknowledgments} }:
It is pleasure to thank Andrew Liddle for useful discussion and comments, and Yun-Song Piao and Yi-Fu Cai for their kind help and discussion.


\vskip 5mm
\begin{thebibliography}{25}

%%%%%%%%***********%%%%%%%%%%%%%%%%***********%%%%%%%%
            % --Inflation Scenario-- %
%%%%%%%%***********%%%%%%%%%%%%%%%%***********%%%%%%%%

\bibitem {KT}
  E. W. Kolb and M. S. Turner, {\it The Early Universe}, Perseus Publishing, 1990, Cambridge, Massachusetts;
  A. Linde, {\it Particle Physics and Inflationary Cosmology}, Harwood Academic, Switzerland.


%%%%%%%%***********%%%%%%%%%%%%%%%%***********%%%%%%%%
            % --WMAP Data-- %
%%%%%%%%***********%%%%%%%%%%%%%%%%***********%%%%%%%%

\bibitem{Spergel}
  D. N. Spergel {\it et al.}, Astrophys. J. Suppl. {\bf 170}, L377 (2007)

\bibitem{Komatsu:2010fb}
  E.~Komatsu {\it et al.},
  %``Seven-Year Wilkinson Microwave Anisotropy Probe (WMAP) Observations:
  %Cosmological Interpretation,''
  arXiv:1001.4538 [astro-ph.CO].
  %%CITATION = ARXIV:1001.4538;%%

\bibitem{Larson:2010gs}
  D.~Larson {\it et al.},
  %``Seven-Year Wilkinson Microwave Anisotropy Probe (WMAP) Observations: Power
  %Spectra and WMAP-Derived Parameters,''
  arXiv:1001.4635 [astro-ph.CO].
  %%CITATION = ARXIV:1001.4635;%%


%%%%%%%%***********%%%%%%%%%%%%%%%%***********%%%%%%%%
            % --Curvature Scenario -- %
%%%%%%%%***********%%%%%%%%%%%%%%%%***********%%%%%%%%

\bibitem{Lyth:2001nq}
  D.~H.~Lyth and D.~Wands,
  %``Generating the curvature perturbation without an inflaton,''
  Phys.\ Lett.\  B {\bf 524}, 5 (2002)
  [arXiv:hep-ph/0110002].
  %%CITATION = PHLTA,B524,5;%%


%\cite{Moroi:2001ct}
\bibitem{Moroi:2001ct}
  T.~Moroi and T.~Takahashi,
  %``Effects of cosmological moduli fields on cosmic microwave background,''
  Phys.\ Lett.\  B {\bf 522}, 215 (2001)
  [Erratum-ibid.\  B {\bf 539}, 303 (2002)]
  [arXiv:hep-ph/0110096].
  %%CITATION = PHLTA,B522,215;%%


%\cite{Dimopoulos:2002kt}
\bibitem{Dimopoulos:2002kt}
  K.~Dimopoulos and D.~H.~Lyth,
  %``Models of inflation liberated by the curvaton hypothesis,''
  Phys.\ Rev.\  D {\bf 69}, 123509 (2004)
  [arXiv:hep-ph/0209180].
  %%CITATION = PHRVA,D69,123509;%%

%\cite{Lyth:2002my}
\bibitem{Lyth:2002my}
  D.~H.~Lyth, C.~Ungarelli and D.~Wands,
  %``The primordial density perturbation in the curvaton scenario,''
  Phys.\ Rev.\  D {\bf 67}, 023503 (2003)
  [arXiv:astro-ph/0208055].
  %%CITATION = PHRVA,D67,023503;%%


%\cite{Lyth:2005ab}
\bibitem{Lyth:2005ab}
  D.~H.~Lyth,
  %``Some recent work on the curvaton paradigm,''
  Nucl.\ Phys.\ Proc.\ Suppl.\  {\bf 148}, 25 (2005).
  %%CITATION = NUPHZ,148,25;%%

%\cite{Lyth:2005fi}
\bibitem{Lyth:2005fi}
  D.~H.~Lyth and Y.~Rodriguez,
  %``The inflationary prediction for primordial non-gaussianity,''
  Phys.\ Rev.\ Lett.\  {\bf 95}, 121302 (2005)
  [arXiv:astro-ph/0504045].
  %%CITATION = PRLTA,95,121302;%%


%%%%%%%%***********%%%%%%%%%%%%%%%%***********%%%%%%%%
        % --Inhomogeneous/Modulated Reheating-- %
%%%%%%%%***********%%%%%%%%%%%%%%%%***********%%%%%%%%

%\cite{Dvali:2003ar}
\bibitem{Dvali:2003ar}
  G.~Dvali, A.~Gruzinov and M.~Zaldarriaga,
  %``Cosmological perturbations from inhomogeneous reheating, freezeout, and
  %mass domination,''
  Phys.\ Rev.\  D {\bf 69}, 083505 (2004)
  [arXiv:astro-ph/0305548].
  %%CITATION = PHRVA,D69,083505;%%

%\cite{Dvali:2003em}
\bibitem{Dvali:2003em}
  G.~Dvali, A.~Gruzinov and M.~Zaldarriaga,
  %``A new mechanism for generating density perturbations from inflation,''
  Phys.\ Rev.\  D {\bf 69}, 023505 (2004)
  [arXiv:astro-ph/0303591].
  %%CITATION = PHRVA,D69,023505;%%

\bibitem{Zaldarriaga:2003my}
  M.~Zaldarriaga,
  %``Non-Gaussianities in models with a varying inflaton decay rate,''
  Phys.\ Rev.\  D {\bf 69}, 043508 (2004)
  [arXiv:astro-ph/0306006].
  %%CITATION = PHRVA,D69,043508;%%

%\cite{Kofman:2003nx}
\bibitem{Kofman:2003nx}
  L.~Kofman,
  %``Probing string theory with modulated cosmological fluctuations,''
  arXiv:astro-ph/0303614.
  %%CITATION = ASTRO-PH/0303614;%%

%\cite{Matarrese:2003tk}
\bibitem{Matarrese:2003tk}
  S.~Matarrese and A.~Riotto,
  %``Large-scale curvature perturbations with spatial and time variations of
  %the inflaton decay rate,''
  JCAP {\bf 0308}, 007 (2003)
  [arXiv:astro-ph/0306416].
  %%CITATION = JCAPA,0308,007;%%

%\cite{Vernizzi:2003vs}
\bibitem{Vernizzi:2003vs}
  F.~Vernizzi,
  %``Cosmological perturbations from varying masses and couplings,''
  Phys.\ Rev.\  D {\bf 69}, 083526 (2004)
  [arXiv:astro-ph/0311167].
  %%CITATION = PHRVA,D69,083526;%%

%\cite{Suyama:2007bg}
\bibitem{Suyama:2007bg}
  T.~Suyama and M.~Yamaguchi,
  %``Non-Gaussianity in the modulated reheating scenario,''
  Phys.\ Rev.\  D {\bf 77}, 023505 (2008)
  [arXiv:0709.2545 [astro-ph]].
  %%CITATION = PHRVA,D77,023505;%%

%\cite{Battefeld:2007st}
\bibitem{Battefeld:2007st}
  T.~Battefeld,
  %``Modulated Perturbations from Instant Preheating after new Ekpyrosis,''
  Phys.\ Rev.\  D {\bf 77}, 063503 (2008)
  [arXiv:0710.2540 [hep-th]].
  %%CITATION = PHRVA,D77,063503;%%

%\cite{Ichikawa:2008ne}
\bibitem{Ichikawa:2008ne}
  K.~Ichikawa, T.~Suyama, T.~Takahashi and M.~Yamaguchi,
  %``Primordial Curvature Fluctuation and Its Non-Gaussianity in Models with
  %Modulated Reheating,''
  Phys.\ Rev.\  D {\bf 78}, 063545 (2008)
  [arXiv:0807.3988 [astro-ph]].
  %%CITATION = ARXIV:0807.3988;%%

%\cite{Kohri:2009ac}
\bibitem{Kohri:2009ac}
  K.~Kohri, D.~H.~Lyth and C.~A.~Valenzuela-Toledo,
  %``On the generation of a non-gaussian curvature perturbation during
  %preheating,''
  JCAP {\bf 1002}, 023 (2010)
  [arXiv:0904.0793 [hep-ph]].
  %%CITATION = JCAPA,1002,023;%%

%%%%%%%%***********%%%%%%%%%%%%%%%%***********%%%%%%%%
            % --DBI involved-- %
%%%%%%%%***********%%%%%%%%%%%%%%%%***********%%%%%%%%

% ************* DBI cosmology earlier works ***************

%\cite{Alishahiha:2004eh}
\bibitem{Alishahiha:2004eh}
  M.~Alishahiha, E.~Silverstein and D.~Tong,
  %``DBI in the sky,''
  Phys.\ Rev.\  D {\bf 70}, 123505 (2004)
  [arXiv:hep-th/0404084].
  %%CITATION = PHRVA,D70,123505;%%

%\cite{Silverstein:2003hf}
\bibitem{Silverstein:2003hf}
  E.~Silverstein and D.~Tong,
  %``Scalar Speed Limits and Cosmology: Acceleration from D-cceleration,''
  Phys.\ Rev.\  D {\bf 70}, 103505 (2004)
  [arXiv:hep-th/0310221].
  %%CITATION = PHRVA,D70,103505;%%

%\cite{Chen:2006nt}
\bibitem{Chen:2006nt}
  X.~Chen, M.~x.~Huang, S.~Kachru and G.~Shiu,
  %``Observational signatures and non-Gaussianities of general single field
  %inflation,''
  JCAP {\bf 0701}, 002 (2007)
  [arXiv:hep-th/0605045].
  %%CITATION = JCAPA,0701,002;%%

%\cite{Easson:2007dh}
\bibitem{Easson:2007dh}
  D.~A.~Easson, R.~Gregory, D.~F.~Mota, G.~Tasinato and I.~Zavala,
  %``Spinflation,''
  JCAP {\bf 0802}, 010 (2008)
  [arXiv:0709.2666 [hep-th]].
  %%CITATION = JCAPA,0802,010;%%

%\cite{Easson:2007fz}
\bibitem{Easson:2007fz}
  D.~A.~Easson, R.~Gregory, G.~Tasinato and I.~Zavala,
  %``Cycling in the throat,''
  JHEP {\bf 0704}, 026 (2007)
  [arXiv:hep-th/0701252].
  %%CITATION = JHEPA,0704,026;%%

%\cite{Gmeiner:2007uw}
\bibitem{Gmeiner:2007uw}
  F.~Gmeiner and C.~D.~White,
  %``DBI Inflation using a One-Parameter Family of Throat Geometries,''
  JCAP {\bf 0802}, 012 (2008)
  [arXiv:0710.2009 [hep-th]].
  %%CITATION = JCAPA,0802,012;%%

%\cite{Arroja:2008yy}
\bibitem{Arroja:2008yy}
  F.~Arroja, S.~Mizuno and K.~Koyama,
  %``Non-gaussianity from the bispectrum in general multiple field inflation,''
  JCAP {\bf 0808}, 015 (2008)
  [arXiv:0806.0619 [astro-ph]].
  %%CITATION = JCAPA,0808,015;%%

% *******END****** DBI cosmology earlier works ********END*******

%\cite{Huang:2007hh}
\bibitem{Huang:2007hh}
  M.~x.~Huang, G.~Shiu and B.~Underwood,
  %``Multifield DBI Inflation and Non-Gaussianities,''
  Phys.\ Rev.\  D {\bf 77}, 023511 (2008)
  [arXiv:0709.3299 [hep-th]].
  %%CITATION = PHRVA,D77,023511;%%

%\cite{Langlois:2008mn}
\bibitem{Langlois:2008mn}
  D.~Langlois and S.~Renaux-Petel,
  %``Perturbations in generalized multi-field inflation,''
  JCAP {\bf 0804}, 017 (2008)
  [arXiv:0801.1085 [hep-th]].
  %%CITATION = JCAPA,0804,017;%%

%\cite{Langlois:2008wt}
\bibitem{Langlois:2008wt}
  D.~Langlois, S.~Renaux-Petel, D.~A.~Steer and T.~Tanaka,
  %``Primordial fluctuations and non-Gaussianities in multi-field DBI
  %inflation,''
  Phys.\ Rev.\ Lett.\  {\bf 101}, 061301 (2008)
  [arXiv:0804.3139 [hep-th]].
  %%CITATION = ARXIV:0804.3139;%%

%\cite{Langlois:2008qf}
\bibitem{Langlois:2008qf}
  D.~Langlois, S.~Renaux-Petel, D.~A.~Steer and T.~Tanaka,
  %``Primordial perturbations and non-Gaussianities in DBI and general
  %multi-field inflation,''
  Phys.\ Rev.\  D {\bf 78}, 063523 (2008)
  [arXiv:0806.0336 [hep-th]].
  %%CITATION = ARXIV:0806.0336;%%

%\cite{Guo:2008sz}
\bibitem{Guo:2008sz}
  Z.~K.~Guo and N.~Ohta,
  %``Cosmological Evolution of Dirac-Born-Infeld Field,''
  JCAP {\bf 0804}, 035 (2008)
  [arXiv:0803.1013 [hep-th]].
  %%CITATION = JCAPA,0804,035;%%


%%%%%%%%%%********** M Y **********%%%%%%%%%%%%%%%%%%%

\bibitem{Li:2008fm}
  S.~Li, Y.~F.~Cai and Y.~S.~Piao,
  %``DBI-Curvaton,''
  Phys.\ Lett.\  B {\bf 671}, 423 (2009)
  [arXiv:0806.2363 [hep-ph]].
  %%CITATION = ARXIV:0806.2363;%%


%%%%%%%%***********%%%%%%%%%%%%%%%%***********%%%%%%%%
        % --Perturbations & non-Gaussianity --%
%%%%%%%%***********%%%%%%%%%%%%%%%%***********%%%%%%%%


%\cite{Maldacena:2002vr}
%\bibitem{M-S-B}
\bibitem{Maldacena:2002vr}
  J.~M.~Maldacena,
  %``Non-Gaussian features of primordial fluctuations in single field inflationary models,''
  JHEP {\bf 0305}, 013 (2003)
  [arXiv:astro-ph/0210603].
  %%CITATION = JHEPA,0305,013;%%

%\cite{Scoccimarro:2003wn}
\bibitem{Scoccimarro:2003wn}
  R.~Scoccimarro, E.~Sefusatti and M.~Zaldarriaga,
  %``Probing Primordial Non-Gaussianity with Large-Scale Structure,''
  Phys.\ Rev.\  D {\bf 69}, 103513 (2004)
  [arXiv:astro-ph/0312286].
  %%CITATION = PHRVA,D69,103513;%%

\bibitem{N.Bartolo:2004uv}
  N.~Bartolo, E.~Komatsu, S.~Matarrese, A.~Riotto
  % `` Non-Gaussianity from inflation: theory and observations ''
  Phys.\ Rept.\  {\bf 402}, 103 (2004)



%%%%%%%********** DBIs: Ref to Mine **********%%%%%%%%%%%%%%
\bibitem{Q.Huang:DBIs}
%\cite{Huang:2008qf}
%\bibitem{Huang:2008qf}
  Q.~G.~Huang,
  %``Spectral Index in Curvaton Scenario,''
  Phys.\ Rev.\  D {\bf 78}, 043515 (2008)
  [arXiv:0807.0050 [hep-th]];
  %%CITATION = ARXIV:0807.0050;%%
%\cite{Huang:2008rj}
%\bibitem{Huang:2008rj}
  Q.~G.~Huang,
  %``N-vaton,''
  JCAP {\bf 0809} (2008) 017
  [arXiv:0807.1567 [hep-th]].
  %%CITATION = ARXIV:0807.1567;%%

\bibitem{Y.Cai:DBIs}
%\cite{Cai:2008if}
%\bibitem{Cai:2008if}
  Y.~F.~Cai and W.~Xue,
  %``N-flation from multiple DBI type actions,''
  Phys.\ Lett.\  B {\bf 680}, 395 (2009)
  [arXiv:0809.4134 [hep-th]];
  %%CITATION = PHLTA,B680,395;%%
%\cite{Cai:2009hw}
%\bibitem{Cai:2009hw}
  Y.~F.~Cai and H.~Y.~Xia,
  %``Inflation with multiple sound speeds: a model of multiple DBI type actions
  %and non-Gaussianities,''
  Phys.\ Lett.\  B {\bf 677}, 226 (2009)
  [arXiv:0904.0062 [hep-th]];
  %%CITATION = PHLTA,B677,226;%%
%\cite{Cai:2010rt}
%\bibitem{Cai:2010rt}
  Y.~F.~Cai and Y.~Wang,
  %``Large Nonlocal Non-Gaussianity from a Curvaton Brane,''
  arXiv:1005.0127 [hep-th].
  %%CITATION = ARXIV:1005.0127;%%

\bibitem{M.Li:DBIs}
%\cite{Li:2008gg}
%\bibitem{Li:2008gg}
 M.~Li and Y.~Wang,
 %``Consistency Relations for Non-Gaussianity,''
 JCAP {\bf 0809}, 018 (2008)
 [arXiv:0807.3058 [hep-th]];
 %%CITATION = ARXIV:0807.3058;%%
%\cite{Li:2008tw}
%\bibitem{Li:2008tw}
  M.~Li and C.~Lin,
  %``Reconstruction of the isocurvaton scenario,''
  arXiv:0807.4352 [astro-ph].
  %%CITATION = ARXIV:0807.4352;%%

%\cite{Wands:2010af}
\bibitem{Wands:2010af}
  D.~Wands,
  %``Local non-Gaussianity from inflation,''
  Class.\ Quant.\ Grav.\  {\bf 27}, 124002 (2010)
  [arXiv:1004.0818 [astro-ph.CO]].
  %%CITATION = CQGRD,27,124002;%%


%%%%%%%%%% ** NEW ** Related to Wrapping/Throat ** ADDED **%%%%%%%%%%%%%%%%%%%%%%%

%%%%%%%%%%% *** ADDED *** %%%%%%%%%%%%%%

%\cite{Chen:2005ad}
\bibitem{Chen:2005ad}
  X.~Chen,
  %``Inflation from warped space,''
  JHEP {\bf 0508}, 045 (2005)
  [arXiv:hep-th/0501184].
  %%CITATION = JHEPA,0508,045;%%

%\cite{Kecskemeti:2006cg}
\bibitem{Kecskemeti:2006cg}
  S.~Kecskemeti, J.~Maiden, G.~Shiu and B.~Underwood,
  %``DBI inflation in the tip region of a warped throat,''
  JHEP {\bf 0609}, 076 (2006)
  [arXiv:hep-th/0605189].
  %%CITATION = JHEPA,0609,076;%%

%\cite{Becker:2007ui}
\bibitem{Becker:2007ui}
  M.~Becker, L.~Leblond and S.~E.~Shandera,
  %``Inflation from Wrapped Branes,''
  Phys.\ Rev.\  D {\bf 76}, 123516 (2007)
  [arXiv:0709.1170 [hep-th]].
  %%CITATION = PHRVA,D76,123516;%%

%% --- Warp & Reheating --- %%

%\cite{Barnaby:2004gg}
\bibitem{Barnaby:2004gg}
  N.~Barnaby, C.~P.~Burgess and J.~M.~Cline,
  %``Warped reheating in brane-antibrane inflation,''
  JCAP {\bf 0504}, 007 (2005)
  [arXiv:hep-th/0412040].
  %%CITATION = JCAPA,0504,007;%%


%\cite{Chialva:2005zy}
\bibitem{Chialva:2005zy}
  D.~Chialva, G.~Shiu and B.~Underwood,
  %``Warped reheating in multi-throat brane inflation,''
  JHEP {\bf 0601}, 014 (2006)
  [arXiv:hep-th/0508229].
  %%CITATION = JHEPA,0601,014;%%

%% -- Warp & Curvaton -- %%
%\cite{Kobayashi:2009cm}
\bibitem{Kobayashi:2009cm}
  T.~Kobayashi and S.~Mukohyama,
  %``Curvatons in Warped Throats,''
  JCAP {\bf 0907}, 032 (2009)
  [arXiv:0905.2835 [hep-th]].
  %%CITATION = JCAPA,0907,032;%%

%\cite{Zhang:2009gw}
\bibitem{Zhang:2009gw}
  J.~Zhang, Y.~F.~Cai and Y.~S.~Piao,
  %``Rotating a Curvaton Brane in a Warped Throat,''
  arXiv:0912.0791 [hep-th].
  %%CITATION = ARXIV:0912.0791;%%

%\cite{Alabidi:2010ba}
\bibitem{Alabidi:2010ba}
  L.~Alabidi, K.~A.~Malik, C.~T.~Byrnes and K.~Y.~Choi,
  %``How the curvaton scenario, modulated reheating and an inhomogeneous end of
  %inflation are related,''
  arXiv:1002.1700 [astro-ph.CO].
  %%CITATION = ARXIV:1002.1700;%%


\end{thebibliography}
\end{document}